\documentclass[aps,twocolumn,superscriptaddress]{revtex4}
\usepackage{times}
\usepackage{color}
\usepackage{float}
\usepackage{epsfig}

\begin{document}

\title{A coevolving model based on preferential
triadic closure for social media networks }
 \affiliation{Department
of Physics, East China Normal University, Shanghai, 200241, P. R.
China}
\author{Menghui Li}
\affiliation{Temasek Laboratories, National University of Singapore, 117508, Singapore}
\affiliation{ Beijing-Hong Kong-Singapore Joint
Centre for Nonlinear \& Complex Systems (Singapore), National
University of Singapore, Kent Ridge, 119260, Singapore}

\author{Hailin Zou}
 \affiliation{ Beijing-Hong Kong-Singapore Joint
Centre for Nonlinear \& Complex Systems (Singapore), National
University of Singapore, Kent Ridge, 119260, Singapore}
\affiliation{Department of Physics, National University of
Singapore, 117542, Singapore}

\author{Shuguang Guan}
\affiliation{Department of Physics, East China Normal University, Shanghai, 200241, P. R. China}

\author{Xiaofeng Gong}
\affiliation{Temasek Laboratories, National University of Singapore, 117508, Singapore}
\affiliation{ Beijing-Hong Kong-Singapore Joint Centre for Nonlinear \& Complex Systems (Singapore), National University of Singapore, Kent Ridge, 119260, Singapore}

\author{Kun Li}
\affiliation{Temasek Laboratories, National University of Singapore, 117508, Singapore}
\affiliation{ Beijing-Hong Kong-Singapore Joint Centre for Nonlinear \& Complex Systems (Singapore), National University of Singapore, Kent Ridge, 119260, Singapore}

\author{Zengru Di}
\affiliation{ School of Systems Science, Beijing Normal University,
Beijing, 100875, P. R. China}

\author{Choy-Heng Lai}
\affiliation{ Beijing-Hong Kong-Singapore Joint
Centre for Nonlinear \& Complex Systems (Singapore), National
University of Singapore, Kent Ridge, 119260, Singapore}
\affiliation{Department of Physics,
National University of Singapore, 117542, Singapore}

\begin{abstract}
The dynamical origin of complex networks, i.e., the underlying
principles governing network evolution, is a crucial issue in
network study. In this paper, by carrying out analysis to the
temporal data of Flickr and Epinions--two typical social media
networks, we found that the dynamical pattern in neighborhood,
especially the formation of triadic links, plays a dominant role
in the evolution of networks. We thus proposed a coevolving
dynamical model for such networks, in which the evolution is
only driven by the local dynamics--the preferential triadic
closure. Numerical experiments verified that the model can
reproduce global properties which are qualitatively
consistent with the empirical observations.\\

\noindent{\bf SUBJECT AREAS:} COMPLEX NETWORKS, STATISTICAL PHYSICS,
THEORETICAL PHYSICS, MODELLING AND THEORY
\end{abstract}
\maketitle

Last decade has witnessed the booming of social media websites, such
as YouTube, Facebook, Second Life, Flickr, Epinions, and Twitter, to
name just a few, which enable users to upload, disseminate, and
share their interesting contents (e.g., photos, videos, or music,
etc.), and even make friends with each other forming online
communities \cite{socialmedia,socialmedia1}. These so called social
media networks, recording the fingerprints of  all participants'
activities, provide prototypes of real networked complex systems. It
is generally believed that the research in this field will enhance
our understandings of the structure of social networks and the
patterns of human behaviors
\cite{Socialscience,CN:REV2002,CN:REV2002-1,CN:REV2003,CN:REV2006,review:humandynamics}.
Moreover, it also has potential for commercial applications
\cite{Marketing}.

Recently, both theoretical and experimental works have been carried
out in analyzing social media networks. For example, Refs.
\cite{Dezso2006} and \cite{Ratkiewicz2010} investigated the dynamics
of network visitation and dynamics of online popularity,
respectively. Ref. \cite{healthbehavior} designed online experiments
to explore how network structure affects the spread of behaviors.
Refs. \cite{Crane2008} and \cite{Wu2007} studied the relaxation
pattern after stimulations in social systems and how attention to
novel items propagates and fades. In addition, the collective
behaviors among online users under social influence have been
studied both theoretically \cite{socialinfluence} and experimentally
\cite{Salganik2006,Lorenz2011}. These investigations have discovered
interesting organization/evolution patterns
\cite{PropagationFlickr,multirelation,Gallos2012}  and correlations
\cite{scalinglaws,correlation} in social media networks.

One of the fundamental issues in the research on networked dynamical
systems is to reveal the possible generic laws governing the
formation/evolution of networks. In order to obtain better
understandings for this issue, the theoretical modelling, especially
based on empirical analysis to real data, is of great importance. In
the seminal work, Barab{\'a}si  and Albert set up a model, known as
BA model after their names later, revealing a general rule in the
growth of scale-free networks: the preferential attachment (PA)
\cite{BAmodel}. Nevertheless, this model \cite{BAmodel} and many
others
\cite{CN:REV2002,CN:REV2002-1,CN:REV2003,CN:REV2006,popularityVSsimilarity}
treat the simplest case of networks with only one type of node and
one type of link inside. In fact, there exist various real-world
complex networks, such as the social media networks, which are
characterized by inherent multiplex nodes and multi-relational
connections. What are the characteristics of the multi-relations in
such systems? How topology and dynamics coevolve? Are there any
interaction patterns between topology and dynamics, which
essentially lead to universal statistical properties of networks? To
obtain deeper insights into these open yet challenging questions in
social media networks, a generic dynamical model with multiplex
nature is desirable.

\begin{figure}
\includegraphics[width=0.9\linewidth]{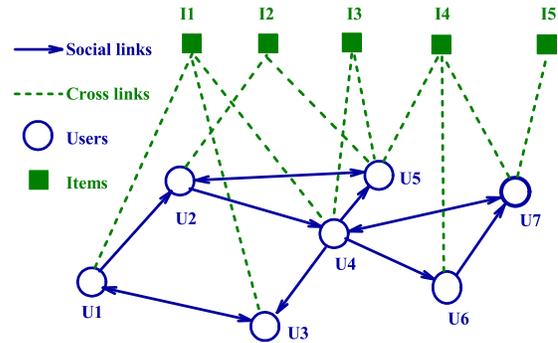}
\caption{Schematic plot of the social media networks, where the
items could be photos (as in Flickr), reviews (as in Epinions),
videos (as in YouTub), etc. In principle, the social links (solid
lines with arrowheads) are directional, while the cross links
(dashed lines) are not. For these two types of links, we can define
four types of degrees (see {\it Methods} for the details). For
example, U4 in the network has indegree ($k_{in}=2$), outdegree
($k_{out}=4$), and favorite degree ($k_f=2$); I4 has popular degree
($k_p=3$). If the directions of social links are ignored, as we will
do in the model, the links among users define the social degree
(which is not the direct summation of indegree and outdegree because
there is overlap between them for a user). In this case, for
example, U4 has social degree ($k_s=5$) and favorite degree
($k_f=2$).}\label{Figure-1}
\end{figure}

In this work, we attempted to set up a model describing the
evolution of social media networks. Our model is based on the
empirical analysis to the real data of Flickr and Epinions, which
are typical social media networks. By analyzing the correlations in
the Flickr and Epinions networks, we found that in such networks,
the exchange of information, the interaction/influence among users,
and the activities of users are essentially within the neighborhood.
It is the local dynamical pattern, especially the formation of
triadic links, that essentially regulates the whole network to
evolve at macroscopic level. Our empirical analysis revealed that
the preferential triadic closure could be one possible dynamical
origin underlying the PA phenomenon in social media networks. Based
on this understanding, we proposed an evolving model starting from
two principles: (1) The evolution of network is governed by local
dynamics, i.e., the preferential triadic closure; (2) The topology
coevolves with the dynamics. This local mechanism in the present
model is different from those based on global information in
previous works \cite{BAmodel,popularityVSsimilarity}.  The validity
of our model has been justified by comparison between the results
from the model and that from the real data. We think that this work
could shed light on the understandings of organization and evolution
in networked dynamical systems, and it also could be helpful in
certain applications, such as designing efficient strategies for
virtual marketing and network marketing, etc.

\noindent\\ \textbf{Results}\\
{\bf Analysis to empirical networks.} As shown in Fig.
\ref{Figure-1}, Flickr and Epinions are typical dual-component
networks which actually represent a broad class of social media
networks consisting of users and items such as photos, videos,
documents, music, blogs and so on (see {\it Methods} for data
description and notations). Due to multiplex nodes and
multi-relations, social media networks are more complicated than the
usual networks involving only one type of node and one type of link.
They are also different from the bipartite networks that are special
dual-component networks previously studied
\cite{mixingmovie,bipartite,bipartite1,bipartite2}. In the
following, we report the main findings of our empirical analysis to
the Flickr and Epinions networks. Our particular attention is paid
to the evolution patterns in these two networks.

\begin{table}
\caption{Exponents $\alpha$ as in  $\kappa (x) \sim x^{\alpha+1}$,
which characterize the cross correlations in the formation of
triadic links and non-triadic links (in the brackets) in the Flickr
and Epinions networks. $\kappa^f$ and $\kappa^{out}$ describe the
relative probabilities of creating new links on the existing
degrees, while $\kappa^{in}$ and $\kappa^{p}$ describe the relative
probabilities acquiring new links on the existing degrees. }
\begin{tabular}{c|c|c|c|c|c}
Networks&$\kappa(x)\setminus x$& $k_f$&  $k_{out}$ & $k_{in}$  & $k_p$ \\
\hline
 &$\kappa^f$&~0.97~[0.70]~&~0.80~[0.36]~&~0.76~[0.33]~&-\\

&$\kappa^{out}$&~0.88~[0.31]~&~0.85~[0.29]~&~0.73~[0.22]~&-\\

\raisebox{1.7ex}[0pt]{Flickr}&$\kappa^{in} $&~0.90~[0.42]~&~0.86~[0.34]~&~0.96~[0.42]~&-\\

&$\kappa^{p}$  & -& -&-&~0.94~[0.94]~ \\ \hline
&$\kappa^f$&~1.24~[0.83]~&~1.17~[0.54]~&~1.13~[0.62]~&-\\

&$\kappa^{out}$&~0.87~[0.30]~&~0.92~[0.45]~&~0.94~[0.46]~&-\\

\raisebox{1.7ex}[0pt]{Epinions}&$\kappa^{in} $&~1.03~[0.52]~&~1.25~[0.75]~&~1.2~[0.76]~&-\\

&$\kappa^{p}$ & -& -&-&~1.06~[0.43]~ \\
\end{tabular}\label{alpha}
\end{table}

\emph{\textbf{Cross correlations.}} Like many other complex
networks, the growth of Flickr and Epinions involves two major
factors: adding new nodes and generating new links. In real
situations, removing nodes and deleting links also happen, but we
neglected them in the present work for simplicity. In particular, we
examined the mechanism forming new links  because this is the
central dynamical process governing network growth. For a growing
network, there are several important questions. The first is: How
new links are formed based on the current status of the network? To
attack this problem, we investigated how generation of new links
depends on the existing degrees in network  (see {\it Methods} for
the definitions of various degrees). Specifically, we extended the
numerical method measuring PA during network growth
\cite{preferentialattachment,measurePA} (see {\it Methods} for more
detail and notations) to calculate the conditional probability, with
which a specific type of degree grows with respect to the existing
degrees.  In this way, the cross correlations among different types
of degrees during network evolution can be identified.

\begin{figure*}
\includegraphics[width=0.9\linewidth]{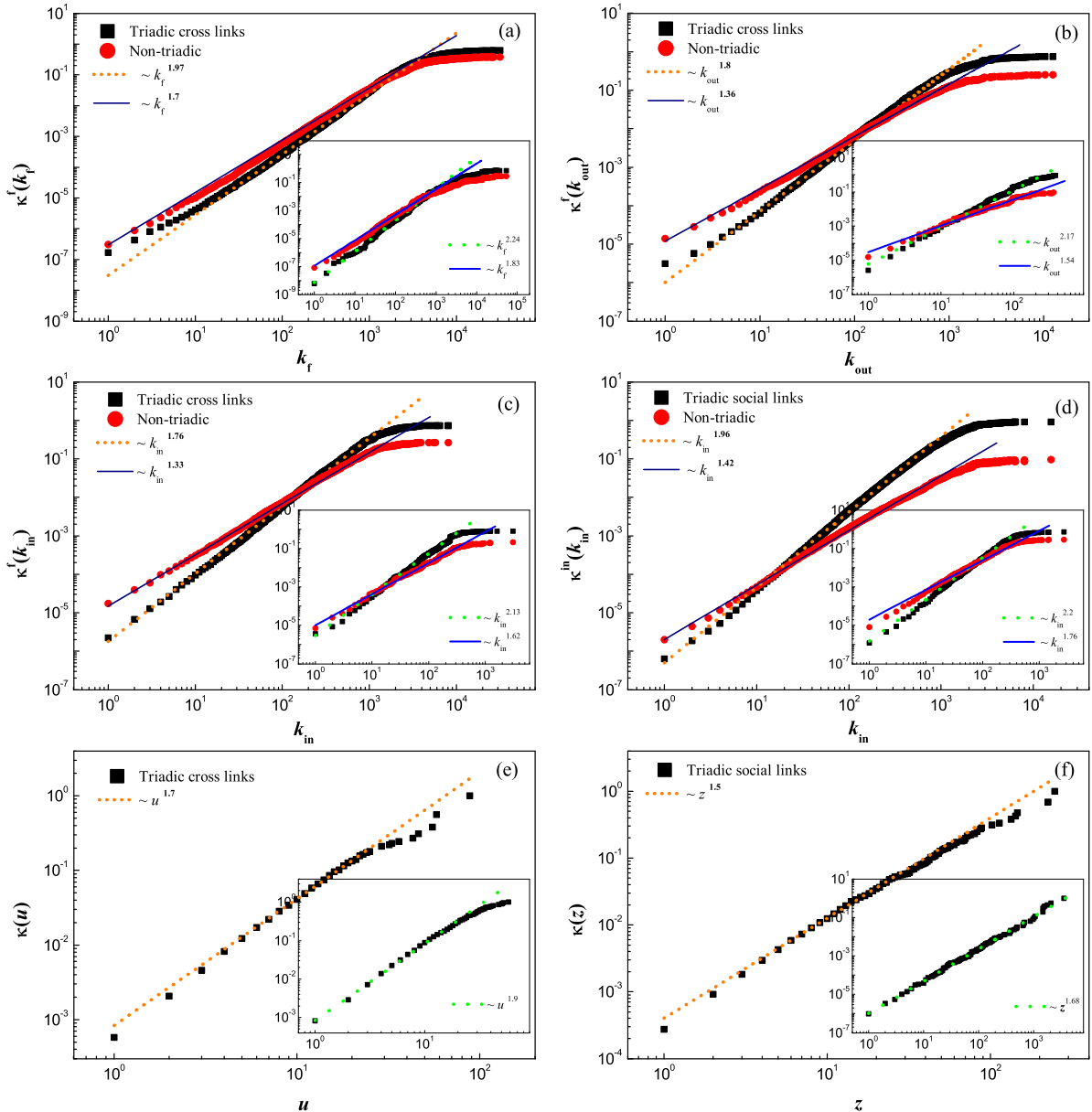}
\caption{The influence of current status on the formation of new
links in the Flickr and Epinions (in the insets) networks. (a)-(d)
The cumulative functions of relative probability, $\kappa^f(k_f)$,
$\kappa^f(k_{out})$,  $\kappa^f(k_{in})$, and $\kappa^{in}(k_{in})$,
respectively, characterizing the cross correlations in the growth of
degrees. For example, in (b), $\kappa^f(k_{out})$ plots the
cumulative probability for a user to build a new cross link given
that his existing outdegree is $k_{out}$. The exponents are obtained
by fitting the curves averaged over different initial $t_0$, and
$\Delta t$ is one day. See the text for the definitions of the
triadic and non-triadic links. (e)-(f) Characterizing the localized
influence among neighboring users. (e) The cumulative functions of
probability for a user to build a cross link to a specific photo
(review) given that $u$ of his neighbors have already
favorite-marked (commented) it. (f) The cumulative functions of
relative probability for a pair of users to build a social link
given that they have already shared $z$ favorite photos (reviews).
The straight lines are guide to the eye through this paper.
}\label{Figure-2}
\end{figure*}

The results of above empirical analysis are illustrated in Figs.
\ref{Figure-2}(a)-\ref{Figure-2}(d) and summarized in Table
\ref{alpha}. From Fig. \ref{Figure-2}(a), we can see that the
relative probability for a user to build a favorite degree is
proportional to the existing favorite degree.  Interestingly, as
shown in Figs. \ref{Figure-2}(b) and \ref{Figure-2}(c), the similar
dependence pattern also exists for the outdegree and indegree.
Moreover, the approximate linear form of the cumulative functions
$\kappa$ in the double-log scale indicates that the relative
probability generating new degree satisfies a power law with respect
to the existing degrees, which can be characterized by the positive
exponent $\alpha$ as in $\kappa (x) \sim x^{\alpha+1}$ with $x$
denoting the degree. Numerical analysis has shown that in the Flickr
and Epinions networks the formation of new links correlates with all
the existing degrees, i.e., the local topological status. In Table
\ref{alpha}, we listed all the characteristic exponents $\alpha$ by
fitting the $\kappa$ functions. They are all positive, between 0 and
1.25, confirming the positive correlation pattern in link growth.

\emph{\textbf{Local dynamics.}} The second question regarding
network growth is: Are the formed links short-ranged or long-ranged
in terms of topological distance? How they depend on the topology?
In other words, do users prefer to generate local connections or
global connections? To attack this problem, we specially divided the
new links into two types: the triadic links and the   non-triadic
links. If a new link can contribute at least one triangle in the
network, it is regarded as a  triadic link. Otherwise, it belongs to
non-triadic link. It is found that most of new links are triadic
links. For example, over $80\%$ social links and over $50\%$ cross
links in Flickr \cite{PropagationFlickr} and over $70\%$ social
links and over $60\%$ cross links in Epinions are triadic links.
These results show that links with short topological distances are
more likely to be established. Moreover, as shown in Figs.
\ref{Figure-2}(a)-\ref{Figure-2}(d) and Table \ref{alpha}, the
formation of both  triadic links and  non-triadic links depends on
the existing degrees with power law relations. However, the
characteristic exponents of the triadic links are significantly
larger than that of the non-triadic links, indicating that the local
topological structure has more severe influence on the formation of
triadic links than on the formation of non-triadic links. This
localized growth pattern might be attributed to the specific rules
in Flickr and Epinions, where a user usually obtains information
from his neighbors and thus has relatively higher probability to
connect to one of his second neighbors (neighbor's neighbors),
either a user or an item.

On the other hand, more importantly, the above empirical analysis
suggests one possible dynamical origin of PA phenomenon: the
preferential triadic closure. As we know, it has been a fundamental
issue to understand the mechanisms underlying PA since the BA model
was successfully proposed \cite{Mossa2002}. In the present work, we
focused on the possible dynamical origin of PA at microscopic level
\cite{preferentialattachment}. In Figs. \ref{Figure-2}(a) and
\ref{Figure-2}(d), we have shown that the formation of triadic links
can approximately lead to  linear PA for degrees $k_f$ and $k_{in}$.
In particular, Fig. \ref{Figure-2}(d) has the exact meaning of PA in
previous studies: The probability to acquire a link is proportional
to its indegree. In fact, as listed in Table I, the diagonal
characteristic exponents $\alpha$ related to triadic links are: 0.97
(1.24), 0.85 (0.92), 0.96 (1.2) and 0.94 (1.06) in the Flickr
(Epinions) network, respectively, which are all close to 1. This
provides an empirical evidence that the formation of triadic links
could be one microscopic dynamical mechanism underlying the linear
PA phenomenon.

\emph{\textbf{Influence within neighborhood.}} The third question
regarding network growth is: How a user's behavior is influenced by
others, especially his neighbors in the network? To answer this
question, we examined the correlations among users' activities
within a neighborhood. For example, we calculated the probability
for a user to favorite-mark (comment) a specific photo (review)
(this will contribute a cross link) given that a certain number of
his neighbors have already favorite-marked (commented) it before he
does. Fig. \ref{Figure-2}(e) plots this relation, where the relative
probability satisfies a power law with positive exponent about 0.7
(0.9) in the Flickr (Epinions) network. This shows that neighboring
users' behaviors of favorite-marking photos (commenting reviews) are
strongly correlated. On the other hand, if two users have already
shared a large number of favorite photos (reviews), it is highly
possible that they have similar appetite or style in photography
(product). Thus it is more likely for them to make friends with each
other due to the common interest. This correlation has been
confirmed in our analysis. As shown in Fig. \ref{Figure-2}(f), the
relative probability for two users, who do not connect to each other
before, to become friends increases with the number of favorite
photos (reviews) that they have commonly shared. The dependence is
also characterized by a power law with exponent about 0.5 (0.68) in
the Flickr (Epinions) network. This provides the evidence that
common interest is in favor of social connections in  these two
networks. The above analysis suggests that in these two networks the
influence or the interplay among users are typically localized in
neighborhood.

\begin{figure*}
\includegraphics[width=\linewidth]{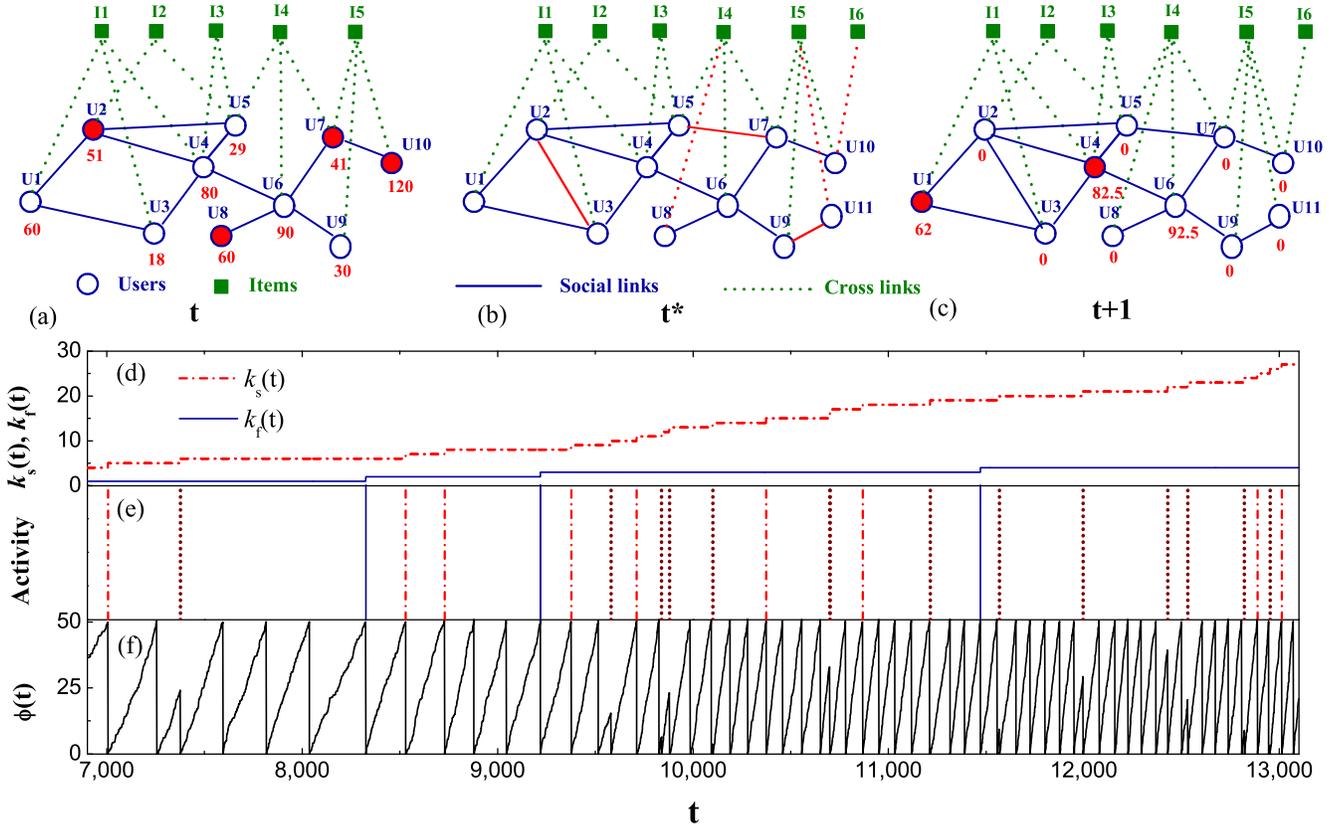}
\caption{Schematic illustration of the coevolution of both topology
(a-c), and dynamical states (d-f) in the model. The numeric tags are
the values of state functions. (a) The network at time $t$, where
some users (solid) are activated according to their states. (b) One
step updating of the network topology, which involves several
activities: New user U11 joins and randomly connects to user U9 and
one of his favorite item I5; New item I6 is created (uploaded) by
activated user U10; The activated users connect to their second
neighbors, including friend of friend (e.g., U2 to U3); favorite
item of friend (e.g., U8 to I4); and the fan of favorite item (e.g.,
U7 to U5). (c) At time $t+1$, the states of users are updated
according to equation (\ref{state}). The states of activated users
at time $t$ are reset to $0$, but some nodes are activated again
according to their states at time $t+1$. (d)-(f) Illustrating the
evolution of  degrees and state function for a specific user during
certain time period in the model. (d) Evolution of the social degree
$k_s(t)$ and the favorite degree $k_f(t)$. (e) The network-involving
activities of the user. The dashed dot lines and the dot lines
indicate the moments when the user initiatively increases his social
degree (e.g., U2 to U3 in (b)),  and the moments when the user
passively increases his social degree (e.g., U5 was connected by U7
in (b)), respectively.  The solid lines represent the moments when
the user increases his favorite degree (e.g., U8 to I4 in (b)). (f)
Evolution of the state function $\phi (t)$. Parameters for the
model:  $m=10, n=100, \mu=0.5, \phi_0=1, \Theta\in[40,4000]$
(uniformly random number), $N=100,000$ (the final size of the
network).}\label{Figure-3}
\end{figure*}

{\bf Modelling.} In this paper we attempted to set up a theoretical
model for social media networks. Our primary target is to
qualitatively reproduce the main properties observed in two
empirical networks. We have two motivations. Firstly, theoretical
modelling is a necessary approach to understand the mechanisms or
generic laws governing the evolution of real-world networks.
Previously, it has been successfully shown that linear PA  can lead
to scale-free property \cite{Connectivity}. Nevertheless, there is a
deeper and interesting question: Is there any microscopic dynamical
origin underlying the phenomenon of PA, which governs the network
growth and leads to various statistical properties in real networks?
This question has attracted much attention previously
\cite{popularityVSsimilarity,Mossa2002,Connectivity,Krapivsky2001,Kleinberg1999},
but it is still worthy of further investigation. Secondly, so far,
although many models have been proposed to describe the evolution of
networked systems
{\cite{CN:REV2002,CN:REV2002-1,CN:REV2003,CN:REV2006}, they mainly
dealt with the networks comprising one type of node
\cite{BAmodel,popularityVSsimilarity} or the bipartite networks
\cite{bipartite,bipartite1,bipartite2}, focusing on the evolution of
network topologies
\cite{CN:REV2002,CN:REV2002-1,CN:REV2003,CN:REV2006,BAmodel,popularityVSsimilarity,bipartite,bipartite1,bipartite2}.
In fact, in realistic systems all the network properties should be
the natural consequence of the coevolution of both dynamics and
topology \cite{adaptive}.

We then based our modelling on the empirical findings reported in
the previous section, i.e., in the Flickr and Epinions networks the
main network-involving activities of users, such as searching,
sending and receiving information, interacting with each other, and
generating new links, are usually limited in the neighborhood.
Especially, new links are more likely to be formed between users and
their second neighbors, namely, the formation of triadic links is
preferential. It is this local dynamical pattern that governs the
evolution of network as a whole. Based on this understanding, we
made the first guiding principle for our modelling:
\begin{itemize}
\item The network evolution should be only driven by local dynamics.
\end{itemize}
For simplicity, we made the following assumptions in our model: (1)
Users and links once join the network, they are never removed; (2)
The social links are symmetric, i.e., their directions are ignored;
(3) New link connecting existing nodes (not the newly added nodes)
is always triadic, i.e., between a user and one of his second
neighbors. In this way, the link growth process can be understood as
a two-step random walk following the information flow in the
network, either via the cross links or via the social links.

In fact, we can consider a model only based on the above principle
to describe the evolution of network topology. For example,
following the third assumption above, we can assign a user certain
probability that is based on the topological information in his
neighborhood to form new links. In our study, we have tried several
ideas of the existing models to simulate the evolution in social
media networks. The results have shown that considering the
topological aspect alone in the modelling will not yield
satisfactory properties in the generated network, e.g., the power
law exponents are usually larger than 2 in these cases, which is
inconsistent with the empirical results in the Flickr and Epinions
networks, where the power law exponents are typically less than 2.
This urged us to consider the fact that in social media networks
dynamics and topology strongly interplay with each other. To obtain
a satisfactory description of such networks, we made the second
guiding principle for our modelling:
\begin{itemize}
\item Both topology and dynamics should coevolve.
\end{itemize}
However, it is not an easy task to formulate the users' behaviors in
real circumstance because they are very complicated due to the
inherent diversity of human dynamics. By carefully examining Flickr,
Epinions, and LinkedIn, etc, we noticed that users'
network-involving activities mainly depend on two factors: the
stimuli from their neighbors and their own initiatives. For example,
in LinkedIn, once a user updates his profile or connects to a new
friend, his friend will automatically receive an email notice from
the system, which inspires him to login the website to act
accordingly. Of course, even without any notice, a user may also
login the website either frequently or occasionally based on his own
habit. To describe such behaviors of users, in our model we
introduced a state function $\phi(i,t)$ for each user, where $i$ is
the user index and $t$ denotes the time. Physically, it represents
the willingness of a user to conduct online activities and it
evolves as:
\begin{equation}\label{state}
\phi(i,t+1)=\phi(i,t) + \mu \sum_j^N
S_{ij}[k_{a}(j,t+1)-k_{a}(j,t)]+ \phi_0 ,
\end{equation}
where $\mu $ and $\phi_0$ are two parameters;
$k_{a}(j,t)=k_{s}(j,t)+k_{f}(j,t)$ is  the sum of social degree and
favorite degree,  i.e., the total degree of user $j$ at time $t$,
which reasonably represents the active level of the user. Note that
in our model we neglected the directions of social links, so there
are only two types of degrees for users, i.e, $k_s$ and $k_f$. In
equation (\ref{state}), the first term at R.H.S. means that the
state function is cumulative. The second term at R.H.S. sums up the
stimuli received by user $i$ from his neighbors. If his neighbors
build new links at time $t$, his state  will be  affected at time
$t+1$, i.e., increasing an amount proportional to the net degree
increase of his neighbors.  Finally, the third term at R.H.S. simply
describes the user's personal initiative by increasing a constant
$\phi_0$ at each time step. To characterize the diversity of users,
we randomly set a threshold, e.g., $\Theta_i \in[40,4000]$, for each
user. If the state function exceeds the threshold, the user will be
activated and has a chance to conduct network-involving events, such
as making friends, favorite-marking photos, or uploading photos in
analogy to Flickr. Once a user is activated or builds a link, his
state $\phi(i,t)$ will be reset to zero at the next time step. The
above modelling, which follows our two guiding principles, though
crude, basically imitates users' logins and activities in the social
media networks in terms of state function.

\begin{figure*}
\includegraphics[width=0.8\linewidth]{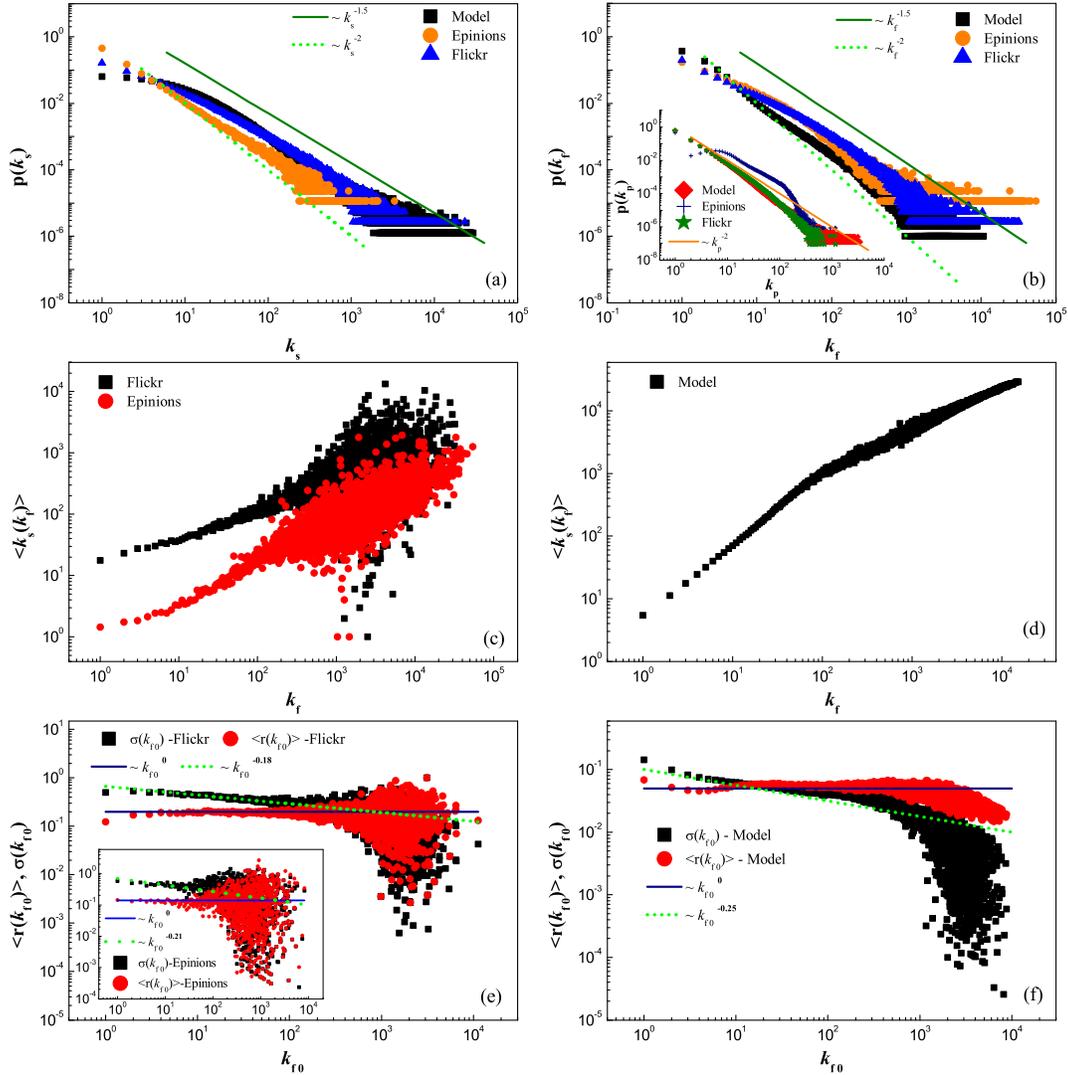}
\caption{Comparing the properties of the model network with that of
the empirical networks. The model parameters are the same as in Fig.
\ref{Figure-3}. Results are averaged over 10 realizations. Numerical
experiments have also been carried out for other parameters, and the
results are qualitatively the same. The empirical networks are at
the final stage in the data collecting window. In the Flickr
network, the total users and photos are about $3.7\times 10^5$ and
$1.1 \times 10^7$, respectively. $\langle k_{out}\rangle=\langle
k_{in}\rangle=50.9$, $\langle k_f \rangle=91.7$, $\langle
k_p\rangle=3.1$. In the Epinions network, the total users and
reviews are about $8.7\times 10^4$ and $1.2 \times 10^6$,
respectively, where $\langle k_{out}\rangle=\langle
k_{in}\rangle=9.46$, $\langle k_f \rangle=159.9$, $\langle
k_p\rangle=11.8$. (a)-(b) The degree distributions. (c)-(d) The
correlations between different types of degrees associated with the
same users in static network. (e)-(f) Characterizing the mean growth
rate r and standard deviation $\sigma$ for favorite degree. For the
model network, $t_0$ = 90,000 and $t_1$ = 100,000. For the empirical
networks, $t_1$ is the final day in the data collecting window, and
$t_0$ is the date about 90 (400) days before $t_1$ for the Flickr
(Epinions) network. }\label{Figure-4}
\end{figure*}

Numerically, the evolution of the model topology obeys the following
rules: (1) At the very beginning, the initial network consists of a
few users ($N_0$) and items ($M_0$), forming a small random network.
The state functions of users evolve according to equation
(\ref{state}). (2) Adding users and items: at every time step, one
new user is added and randomly connects a user and one of his
favorite items. In the mean time $m$ users are randomly selected
from the activated users, and each of them creates one new item. (3)
Adding links: at each time step, $n$ users are randomly selected
from the activated users, and each of them connects to one of his
second neighbors (either user or item) via a two-step random walk if
they do not connect each other before. In (2) and (3), if the number
of activated users is less than $m$ or $n$, respectively, the
insufficient part can be randomly chosen to complement. A schematic
illustration of the above procedure is plotted in Fig.
\ref{Figure-3}, where the state and the topology of the network are
shown to coevolve for one step driven by the local dynamics. As
shown in Figs. \ref{Figure-3}(d)-\ref{Figure-3}(f), with the
increasing of social degree, the period of the state $\phi(t)$ for a
user could become smaller, which means that those users with larger
social degrees are more likely to participate in network-relating
events.

{\bf Numerical verifications.} We carried out numerical simulations
to validate the model. It has been shown that the model can
reproduce the main properties observed in the two empirical
networks, for example, the highly skewed distributions of degrees,
the long-term temporal correlation, the cross correlations among
degrees, the correlation pattern among users' behaviors, etc. In the
following, we compared the results of the model with the empirical
analysis to the Flickr and Epinions networks.

\emph{\textbf{Degree distributions.}} For a complex network, the
degree distribution is one of the most important statistical
properties. Figs. \ref{Figure-4}(a) and \ref{Figure-4}(b) compare
the degree distributions of the model with that of the empirical
networks. In both cases, the distributions were calculated for the
static networks at the final stage. It is shown that all the three
types of degrees approximately exhibit power-law scaling, and the
distributions in the model are qualitatively consistent with the
counterparts of the  empirical networks. In particular, the power
law exponents for the social degrees and the favorite degree can be
less than $2$ in certain parameter regimes in the model, which are
consistent with the empirical observations in many online social
networks \cite{Orkut,tianya,digg}. If we do not consider the
coevolution of dynamics and topology in the model, this property
cannot be reproduced.

\emph{\textbf{Correlations in static network.}} For a network
inherently characterized by multi-relation links, it is natural to
investigate the correlations among different degrees associated with
the same nodes. To this end, we calculated the Pearson's correlation
coefficients (PCC) between different pairs of degrees at the same
nodes. In Fig. \ref{Figure-4}(c) the correlation between the social
degree and the favorite degree is plotted for the empirical
networks, where the PCC is about 0.35 (0.69) for the Flickr
(Epinions) network, showing that they are correlated to some extent.
In Fig. \ref{Figure-4}(d), the same correlation pattern is found for
the model, where the PCC is about 0.90. The quantitative deviation
for the PCCs has the following reason: In real system, the formation
of cross links are affected by many factors besides the social
degree of users, and therefore the PCC is relatively small; In our
model, we only considered the influence of social degree on the
formation of cross links, so the PCC is larger than that in
empirical networks. We have computed all the pairwise PCCs among the
four types of degrees of users in the empirical networks. They are:
0.35 ($k_s$ vs $k_f$), 0.32 ($k_{out}$ vs $k_f$), 0.4 ($k_{in}$ vs
$k_f$), and 0.76 ($k_{in}$ vs $k_{out}$) for the Flickr; and
accordingly 0.69, 0.33, 0.82, and 0.46 for the Epinions. These
results confirm that in the two empirical networks, different types
of degrees of users are all positively correlated with each other.
Our model exhibits qualitatively consistent results. We noticed that
some PCCs in the empirical networks are relatively small. They only
show that different types of degrees in the network correlate to
each other to some extent, but one type of degree cannot fully
characterize the other.

\begin{table}
\caption{Exponents $\beta(r)$ and $\beta(\sigma)$ as in $r(k_0) \sim
k_0^{-\beta(r)}$ and $\sigma(k_0) \sim k_0^{-\beta(\sigma)}$, which
characterize the mean degree growth rate $r$ and the standard
deviation $\sigma$ in the evolution of the model network and the
empirical networks ( Flickr before Epinions in the parentheses). The
parameters are the same as in Fig. \ref{Figure-3} and Fig.
\ref{Figure-4}.}
\begin{tabular}{c|c|c|c}
       &~~~~$k_f$~~~~ & ~~~~$k_{s}$~~~~ & ~~~~$k_{p}$ ~~~~\\
\hline
$\beta(r)$ & ~0.0~(0.0 \& 0.0)~&~0.05~(0.10 \& 0.0)~&~0.0~(0.0 \& 0.0) \\

$\beta(\sigma)$&0.25~(0.18 \& 0.21)~&~0.28~(0.23 \& 0.18)~&~0.5~(0.17 \& 0.2)\\
\end{tabular}\label{Gibrat}
\end{table}

\emph{\textbf{Temporal pattern in degree growth.}} Flickr and
Epinions turn out to be very successful commercially. They keep
expanding rapidly since foundation. From the viewpoint of network,
it is important to investigate the temporal growth pattern of
degrees. We considered the growth rate $r$ of a specific type of
degree \cite{scalinglaws}: $ r \equiv \ln {k_1}/{k_0}$, where $k$
could be any type of degree in the empirical networks. $k_0=k(t_0)$
and $k_1=k(t_1)$ are degrees at time $t_0$ and $t_1$, respectively.
By keeping watching on those nodes with degree $k_0$ (at time $t_0$)
during period $(t_0,t_1)$, we can calculate the average conditional
growth rate $\langle r(k_0)\rangle$, and the standard deviation
$\sigma(k_0)$. Examples are shown in Fig. \ref{Figure-4}(e), where
the growth rate of $k_f$ is almost independent of initial degree
$k_{f0}$. Remarkably, its deviation approximately satisfies a power
law scaling: $\sigma(k_{f0}) \sim k_{f0}^{-\beta}$, known as the
generalized Gibrat's law in economic systems. Particularly, $\beta<
1/2$ indicates the nontrivial long-term correlation pattern
\cite{scalinglaws}. In Fig. \ref{Figure-4}(f), for instance, we
characterized the growth pattern for degree $k_f$ in the model, and
found that it is well consistent with Fig. \ref{Figure-4}(e). We
further confirmed that this long-term correlation pattern exists in
the growth of all types of degrees in the Flickr  and Epinions
networks. In Table \ref{Gibrat}, we computed the characteristic
exponents $\beta(r)$ and $\beta(\sigma)$ for three types of degrees,
i.e., $k_s$, $k_f$ and $k_p$. Note that in our model only three
types of degrees are defined because the directions of social links
are neglected. It is found that all the exponents $\beta(\sigma)$
are less than $1/2$, as observed in economic systems
\cite{economics1,economics2,economics3,economics4} and online
communication networks \cite{scalinglaws}. For comparison, the
corresponding characteristic exponents for the model network are
also listed in Table \ref{Gibrat}. It is seen that for $k_f$ and
$k_s$, $\beta(\sigma)$=0.25 and 0.28, respectively, which are well
consistent with the empirical results where $\beta(\sigma)$=0.18 and
0.23 in Flickr, as well as $\beta(\sigma)$=0.21 and 0.18 in
Epinions, respectively. In our model, for simplicity, we do not
define state functions for items, so the increase of $k_p$ is
basically of random nature. As a consequence, the corresponding
$\beta(\sigma)$ is approximately 0.5. We emphasize that in the model
the values of $\beta(\sigma)$ can be influenced by the heterogeneity
of threshold parameter $\Theta$. For example, if $\Theta$ is set as
a constant in the model, $\beta(\sigma)$ turns out to be close to
0.5, showing very weak temporal correlation. This, on the other
hand, suggests that the diversity of users' behaviors might be one
reason for the long-term correlation during network evolution.

\emph{\textbf{Cross correlations in the formation of degrees.}} In
previous empirical analysis, we have found that in the Flikcr and
Epinions networks the probability to build a new link depends  on
the existing degrees, as shown in Figs.
\ref{Figure-2}(a)-\ref{Figure-2}(d) and  Table \ref{alpha}. To
verify our model, we applied the same analysis and summarized the
results in Table \ref{alpha2}. It is seen that the characteristic
exponents $\alpha$ for the model are qualitatively consistent with
that for the empirical networks.

\begin{table}
\caption{Exponents $\alpha$ for the model and the empirical networks
(Flickr before Epinions in the parentheses). The parameters are the
same as in Fig. \ref{Figure-3} and Fig. \ref{Figure-4}. }
\begin{tabular}{c|c|c|c}
 $\kappa(x)\setminus x$& ~~~~$k_f$~~~~& ~~~~$k_{s}$~~~~&~~~~$k_p$~~~~\\
\hline
$\kappa^f$    & 0.95~(0.97 \&1.24) & 0.84~(0.79 \& 1.16)  &-\\
$\kappa^{s}$  & 0.85~(0.90 \& 0.95) & 0.89~(0.84 \&1.10)   &-\\
$\kappa^{p}$    & -& - & 0.91~(0.94 \&1.06) \\
\end{tabular}\label{alpha2}
\end{table}

\emph{\textbf{Local interaction pattern.}} In previous empirical
analysis, we have revealed that the users' behaviors correlate with
each other in the neighborhood, as shown in Figs. \ref{Figure-2}(e)
and \ref{Figure-2}(f). It is found that the model can reproduce
similar correlation patterns for neighboring users, as shown in
Figs. \ref{Figure-5}(a) and \ref{Figure-5}(b). Since Flickr and
Epinions have two types of nodes, it is also important to
investigate the correlation between neighboring users and items. To
this end, we calculated the average  nearest neighbors' degree
\cite{correlation}. Here the term ``neighbors'' specially refers to
the relation between users and their connecting items. We defined
two quantities: the average popular degree of user $i$'s favorite
items and the average favorite degree of item $\lambda$'s fans as
\begin{equation}
k_p^{nn}(i)=\sum_{\lambda}^M (C_{i\lambda} \sum_j^N C_{j\lambda})/
\sum_{\lambda}^M C_{i\lambda},
\end{equation}
\begin{equation}
 k_f^{nn}(\lambda)=\sum_{i}^N (C_{i\lambda}\sum_{\rho}^M C_{i\rho})/\sum_i^N C_{i\lambda}.
\end{equation}
Averaging the above two quantities over the whole network gives two
functions: $\langle k_p^{nn}(k_f)\rangle$ and $\langle
k_f^{nn}(k_p)\rangle$, where the superscript $nn$ denotes ``nearest
neighbor''. To some extent, they reflect the pattern in users'
behaviors in the process of favorite-marking photos (or commenting
reviews). Interestingly, it is found that these two quantities
exhibit different patterns as shown in Fig. \ref{Figure-5}(c).
$\langle k_f^{nn}(k_p)\rangle$ positively correlates with $k_p$ when
it is small (e.g., $ k_p<10$ in Flickr), but the correlation becomes
negative when $k_p$ is large. This shows that in the Flickr
(Epinions) network, the popular photos (reviews) (with large $k_p$)
are most favorite-marked (commented) by common users (with small
$k_f$). Meanwhile, it is also found that $\langle
k_p^{nn}(k_f)\rangle$ approximately keeps constant, indicating that
users in the Flickr (Epinions) network  do not seem to care about
the popularity of photos (reviews) when they favorite-mark (comment)
them. This result differs from that in bipartite networks, e.g., the
user-movie network, where $\langle k_p^{nn}(k_f)\rangle$ is shown to
be negatively correlated with $k_f$ \cite{mixingmovie}. This
difference may be due to the different spreading modes of photos
(reviews) and movies. For instance, in the user-movie network, a
popular Hollywood blockbuster is just like global information that
everyone knows, but in the user-photo or user-review networks there
is no such counterpart. Instead, in Flickr (Epinions) users mainly
favorite-mark photos (comment reviews) based on localized
information. In Fig. \ref{Figure-5}(d), we computed the above two
quantities for the model, and found that the model can present
similar correlation patterns to the empirical analysis. Moreover, in
our model we found the heterogeneity of  threshold parameter
$\Theta$ can also  affect the nearest neighbor  correlations.

\begin{figure*}
\includegraphics[width=0.8\linewidth]{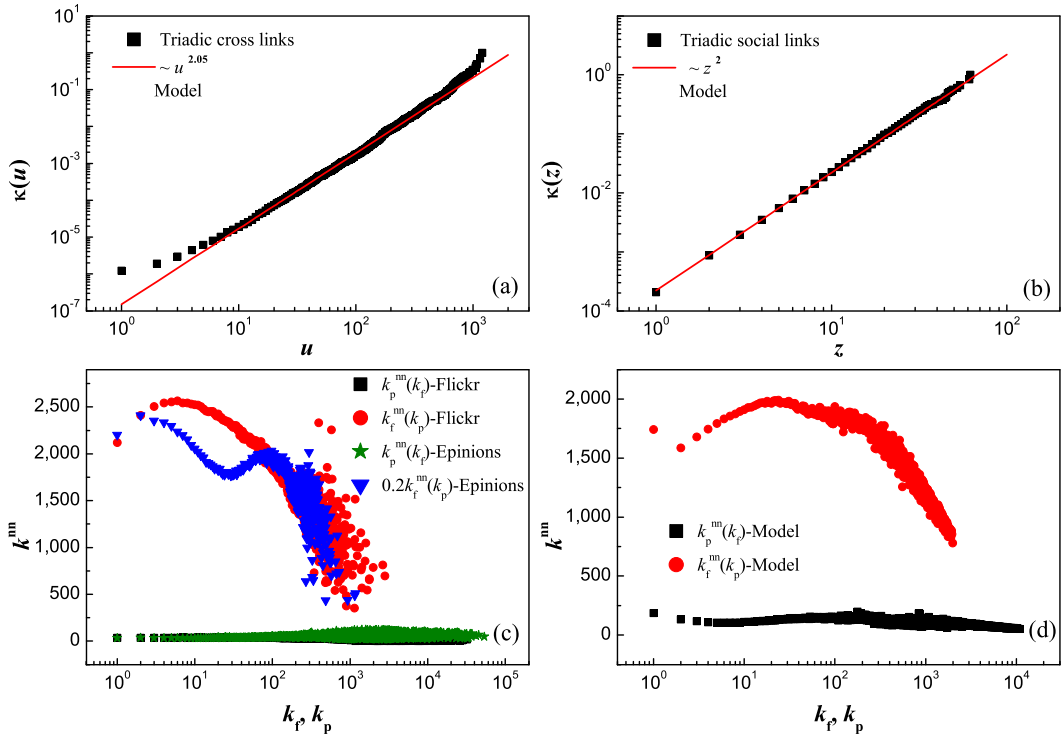}
\caption{Comparing the properties of the model network with that of
the empirical networks. (a)-(b) Characterizing the local influence
patterns among users in the model, as compared with Figs.
\ref{Figure-2}(e) and \ref{Figure-2}(f). (c)-(d) Characterizing the
correlations between neighboring users and items in the model and in
the empirical networks, i.e., $\langle k_p^{nn}(k_f)\rangle$ and
$\langle k_p^{nn}(k_f)\rangle$. The parameters for the model are the
same as in Fig. \ref{Figure-3}.}\label{Figure-5}
\end{figure*}

\noindent\\
\textbf{Discussion}\\
In this paper, we first carried out empirical analysis to the Flickr
and Epinions networks. Our study revealed both temporal  and
topological correlation patterns. Especially, it is found that the
network growth is essentially governed by the preferential formation
of triadic closures. Motivated by the empirical findings, we
proposed a coevolving dynamical model that starts from local
dynamics, and is able to qualitatively reproduce the main properties
observed in the empirical networks. Particularly, to our knowledge,
this is the model based on local information which can exhibit: (1)
Power law degree distributions with exponents less than 2; (2)
Long-term growth pattern of degrees; (3) The nearest neighbors
correlations that are consistent with real networks. Based on both
empirical analysis and theoretical modelling, we think that the
preferential formation of triadic closure could be one dynamical
origin governing the evolution in a broad class of social media
networks.

We emphasize two major characteristics of the model. First, in the
present model, users preferentially generate triadic links based on
their state functions, which only involves localized information and
interaction. This differs from many previous models, which
implicitly require global information
\cite{BAmodel,popularityVSsimilarity}. Second, in our model, both
dynamics and topology coevolve, which naturally leads to a dynamical
network with various correlation patterns. This is also different
from many previous models  that only considered the evolution of
network topology. Due to the ignorance of dynamics, they usually
failed to exhibit correlation properties in networks, for example,
the long-term correlation in degree growth \cite{scalinglaws}.  Of
course, the preferential triadic closure is only one possible
mechanism underlying linear PA phenomenon, which can lead to
scale-free properties. Due to the diversity and different nature of
complex networks, there might be other microscopic mechanisms
governing the evolution of such networks.

It should be pointed out that our model can only reproduce
qualitatively consistent results compared with the empirical
analysis to real data. We attribute the quantitative mismatch to the
simplifications made to set up the model. For example, in our model
users are limited to obtain information only from their
neighborhoods. Actually, real situations could be much more
complicated. Research has shown that although over 80\% users make
their new friends in their second neighbors and over 50\% of users
find their favorite photos from their friends in Flickr
\cite{PropagationFlickr}, users can access photos through various
other channels, including those far from their neighborhoods. For
example, by viewing the top-ranking photos or the most downloaded
photos provided by the website, users may accordingly build links
with large topological distance. These factors will lead to both
long distance correlation and multiscale properties in the network.
Especially they will affect the statistics of the small degrees. All
these problems deserve further investigations in the future.

\noindent\\\textbf{Methods}\\
{\bf Data description and notations.} Flickr was founded in 2004. As
the most famous website sharing photos, currently it has millions of
active users and billions of photos. In Flickr, users basically
involve three activities: uploading photos, favorite-marking photos
and making friends. A user can claim any interesting photos as
favorites (called \emph{favorites} in Flickr). Once he
favorite-marks a photo, he will automatically be a fan of this photo
and thus in its fan list. Then he can retrieve the information of
users in the fan list if he likes. Similarly, a user can
unilaterally declare any other users as friends (called
\emph{contacts} in Flickr). For example, when user $i$ declares user
$j$ as a friend, user $j$ appears in the friends list of user $i$,
and his profile information, such as his favorite photos and
friends, is also available to user $i$.

The data set for our study is collected by daily crawling Flickr
over 2.5 million users from Nov 2, 2006 to Dec 3, 2006, and again
daily from Feb 3, 2007 to May 18, 2007. Totally, there are 104 days
in the time window for data collection
\cite{PropagationFlickr,GrowthFlickr}
(http://socialnetworks.mpi-sws.org/). During this period, over 9.7
million new social links are formed, and over 950 thousand new users
are observed. In particular, all temporal information for uploaded
photos is stamped, including when the owner of a photo uploaded it,
and  who (and when) marked the photo as favorites, etc. For the
purpose of network analysis, here we only considered the users who
at least have one favorite photo and one friend.  With this
constraint, there are about 370 thousand users and 11.1 million
photos in the data. Moreover, we assumed that the uploaders have
marked their photos as favorite by default.

For analysis purpose, we first mapped the data into a network
(referred to as the Flickr network throughout the paper), which is
characterized by the dual components and the dual links, as
schematically shown in Fig. \ref{Figure-1}. It has two types of
nodes, i.e., $N$ users and $M$ items (photos) totally. Meanwhile,
there are also two types of links, i.e., the links among users and
the links connecting users and photos. We call them as the social
links and the cross links, respectively. In the Flickr network,
information can flow either via the social links or the cross links.
Note that, in principle, the social links in the Flickr network are
directional.

Mathematically, we can use two matrices to characterize the topology
shown in Fig. \ref{Figure-1}. $S$ is an $N \times N$ adjacency
matrix representing the social links among users, with element
$S_{ij} = 1$ if user $i$ declares user $j$ as his friend, otherwise
$0$. Note that $S$ is asymmetrical. Similarly, $C$ is an $N \times
M$ adjacency matrix characterizing the cross links, with element
$C_{i\lambda}=1$ if user $i$  favorite-marks photo $\lambda$,
otherwise $0$. To be specific, we defined the following types of
degrees to characterize the multi-relational connections in the
Flickr network. Two degrees are related to the cross links: (1) the
favorite degree: $k_f(i)=\sum_{\lambda}C_{i\lambda}$, i.e., the
number of favorite photos marked by user $i$, and (2) the popular
degree $k_p(\lambda)=\sum_{i}C_{i\lambda}$, i.e., the number of fans
for photo $\lambda$, which reasonably represents its  popularity
extent in the network. Actually, $k_f$ and $k_p$ are two
perspectives of the cross links connecting users and photos. In
addition, since the social links in the network are directional, we
accordingly defined another two types of degrees as: (3) the
indegree: $k_{in}(j)=\sum_{i}S_{ij}$, i.e., the number of users who
claim user $j$ as friend, and (4) the outdegree,
$k_{out}(i)=\sum_{j}S_{ij}$, i.e., the number of friends claimed by
user $i$. Physically, outdegree and favorite degree together reflect
the active level of a user, while the indegree represents the impact
of the user. If we do not discriminate the directions of the social
links, there are three types of degrees in the network: the popular
degree ($k_p$), the favorite degree ($k_f$), and the social degree
($k_s$) that is the number of friends for a user.

Epinions is a product review website established in 1999. In
Epinions, users also basically involve three activities: writing
reviews about products, commenting reviews and expressing their
trust or distrust to other users. Once a user comments a review, he
will automatically be in its commenting list. A user can
unilaterally declare any other users as trust users or distrust
users if he thinks their reviews to be valuable or inaccurate. Trust
relationships are publicly accessible in Epinions but not the
distrust relationships. The data set of Epinions for our study
contains product reviews and review ratings before May 30, 2002, and
both trust and distrust relationships before August 12, 2003
\cite{Epinions}
(http://www.trustlet.org/wiki/Extended\_Epinions\_dataset). In
particular, all temporal information is available. For the purpose
of network analysis, here we only consider the users who at least
have one review and one trust relation before May 30, 2002. With
this constraint, there are 87,577 users, 546,883 trust relations,
1,198,115 reviews and 13,668,319 comments. Moreover, we assumed that
the review writers have commented their own reviews by default.
Apparently, the users, the reviews, trusting users, and commenting
reviews in Epinions, just correspond to the users, the photos,
making friends, favorite-marking photos in Flickr, respectively.
Therefore, these two networks are topologically equivalent. For
simplicity and convenience, we did not define another set of
notations for the Epinions network. Throughout this paper, we used
the same notations defined above for both networks. When necessary,
we pointed out their physical implications in their own contexts.
For example, in the Epinions network, the popular degree $k_p$ and
the favorite degree $k_f$ specifically refer to the number of users
commenting a review and the number of comments given by a user,
respectively.

{\bf Measuring preferential attachment.} In Refs.
\cite{preferentialattachment,measurePA}, a numerical method is used
to measure the preferential attachment (PA) growth of network. Given
knowing the temporal order in which the nodes join the network, the
essential idea of the method is to monitor to which existing node
new nodes connect, as a function of the degree of the old node. We
can extend this method to characterize the multi-relational
correlations among various degrees in the social media networks such
as Flickr and Epinions. We take an example to briefly explain the
method as follows: (1) At time $t_0$, we mark the nodes with
$k_{out}$ outdegree as ``$t_0$ nodes", denoting their number as
$C(k_{out})$. (2) After the evolution of a period $\Delta t$, the
favorite degrees of the ``$t_0$ nodes" have increased due to the
evolution of the network (of course, the other types of degrees also
change). We count the favorite degree acquired by the ``$t_0$ nodes"
as $A(k_{out})$. Since we divided the newly generated links into two
types, i.e., the triadic and non-triadic, we have
$A(k_{out})=A_T(k_{out})+A_N(k_{out})$, where the subscripts $T$ and
$N$ denote the two types, respectively.  (3) The histogram providing
the number of favorite degree acquired by the ``$t_0$ nodes" with
exact $k_{out}$ outdegree, after normalization, defines a function
\cite{measurePA}:
\begin{equation}
%\Pi_i^f(k_{out})
%=\frac{\frac{A_i(k_{out})}{C(k_{out})}}{\sum_{k_{out}'}\frac{A(k_{out}')}{C(k_{out}')}},\label{limh}
\Pi_i^f(k_{out})
=\frac{A_i(k_{out})}{C(k_{out})}/\sum_{k_{out}'}\frac{A(k_{out}')}{C(k_{out}')}, \label{limh}
\end{equation}
where, $i$ can be either $T$ or $N$. It has been proven that if PA
mechanism exists, the conditional probability with which the
favorite degree grows with respect to the existing outdegree follows
a power law:
\begin{equation}
\Pi_i^f(k_{out})\propto k_{out}^{\alpha}.
\end{equation}
Numerically, it is convenient to examine the cumulative function of
$\Pi_i^{f}(k_{out})$,  which will also follow a power law, i.e.,
\begin{equation}
\kappa_i^f(k_{out})=\int^{k_{out}}_{0} \Pi_i^f(k'_{out})dk'_{out}
 \propto k_{out}^{\alpha + 1}.
\label{cumulative}
\end{equation}
Similarly, the above calculation can be applied to any pair of
degrees for $k_{in}$, $k_{out}$, $k_f$, and $k_p$.

\noindent\\ \textbf{Acknowledgments} \\
We thank A. Mislove for sharing the Flickr database, and P. Massa
for sharing the Epinions database. SGG is sponsored by the following
funding agencies: Science and Technology Commission of Shanghai
Municipality under grant No. 10PJ1403300; Innovation Program of
Shanghai Municipal Education Commission under grant No. 12ZZ043; and
the NSFC under grant No. 11075056 and 11135001. ZRD is supported by
NSFC under Grant No. 61174150. This work is supported by Temasek
Laboratories at National University of Singapore through the DSTA
Project No. POD0613356.

\noindent \\ \textbf{Author contributions} \\
M.H.L., H.L.Z., S.G.G., X.F.G., Z.R.D. and C.H.L. designed research
and analyzed the data of Flickr; K.L. analyzed the data of Epinions;
M.H.L. and S.G.G. performed research and wrote the paper. All
authors reviewed and approved the manuscript.

\noindent \\ \textbf{Additional information}\\
 Competing financial interests: The authors declare no competing financial interests.
\noindent \\
\\
Correspondence should be addressed to S. G. Guan
(guanshuguang@hotmail.com).

\end{document}